\documentclass[aps,prd,floats,floatfix,nofootinbib,twocolumn]{revtex4-1}
\usepackage{graphicx,graphics,epsfig}
\usepackage{amssymb}
\usepackage{epstopdf}
\usepackage{hyperref}

\usepackage{amsmath}
\usepackage{bbold}
\usepackage{color}

\usepackage{graphicx,graphics,epsfig}

\begin{document}

\title{The Dilaton-like Higgs boson with scalar singlet dark matter}

\author{Robyn Campbell}
\email{RobynCampbell@cmail.carleton.ca}

\author{Stephen Godfrey}
\email{godfrey@physics.carleton.ca}

\author{Alejandro de la Puente }
\email{apuente@physics.carleton.ca}

\affiliation{Ottawa-Carleton Institute for Physics, Carleton University, 1125 Colonel By Drive, Ottawa, Ontario K1S 5B6, Canada}

\date{July 7, 2016}                                  % Activate to display a given date or no date

\begin{abstract}
We study a model with a Higgs-like dilaton and a 
Standard Model gauge-singlet scalar dark matter candidate.  We begin by
updating the status of identifying the observed 125 GeV Higgs-like boson with the pseudo 
Nambu-Goldstone boson that arises from the spontaneous breaking of scale invariance using 
recent Higgs boson signal strength measurements by the ATLAS and CMS collaborations. We 
then constrain the extended model with recent constraints on the Higgs invisible width,
the observed dark matter relic abundance and the latest dark matter direct detection limits. 
We found that the magnitude of the dilaton-$\gamma\gamma$ and dilaton-glue-glue coupling is
constrained to be close to the standard model values.  The mass of the dark matter candidate
is contrained to be greater than half the dilaton mass by relic abundance limits and Higgs
invisible width limits.  Dark matter direct detection limits allow only small mass regions which
will be further constrained by upcoming DEAP measurements.

\end{abstract}

\maketitle 

\section{Introduction}

Little is known about what lies behind the mechanism of Electroweak Symmetry Breaking (EWSB). Within the Standard Model 
EWSB is realized with an elementary scalar field and a negative mass term in the scalar potential. 
The negative mass term induces an instability that causes the Higgs field to condense leading to 
spontaneous symmetry breaking of the SU(3)$_{\text{c}}$$\times$SU(2)$_{\text{W}}$$\times$U(1)$_{\text{Y}}$ gauge 
symmetry down to SU(3)$_{\text{c}}$$\times$U(1)$_{\text{em}}$. The scale at which the symmetry 
breaks and the spectrum depends strongly on the size of the Higgs mass parameter. However, 
this parameter is strongly sensitive to high energy scales; if the Standard Model is 
the correct theory up to the Planck Scale, the Higgs mass parameter would receive very 
large quantum corrections. To keep the Higgs mass parameter close to the electroweak scale, 
one will have to fine-tune the bare Higgs mass parameter order by order in perturbation theory  
for the physical Higgs mass to be $125$ GeV~\cite{Aad:2012tfa,Chatrchyan:2012ufa,Khachatryan:2016vau}. 
This naturalness problem has led many to believe that the absence of very large quantum 
corrections can only be the result of additional symmetries protecting the Higgs mass parameter. 
One example is supersymmetry, where quadratically divergent contributions to the Higgs mass from 
top quark loops are canceled by loops of scalar particles with the same gauge quantum 
numbers as the top quark (See~\cite{Martin:1997ns} and references therein). However, supersymmetry cannot be an exact symmetry and thus a natural resolution of the naturalness problem would require a supersymmetry breaking scale that is not too large~\cite{Dimopoulos:1995mi}.

Another possibility pursued in a number of papers is that the properties of the Standard Model (SM)
Higgs boson are mainly fixed by the approximate conformal invariance in the limit when the Higgs 
potential is turned off.  In this case the Higgs vacuum expectation value (vev) will spontaneously break the 
approximate conformal symmetry as well as the EWS.  In this scenario the Higgs is identified
with the dilaton which is the Goldstone boson associated with the spontaneously 
broken  conformal symmetry  
and the Higgs couplings are mainly dictated by this conformal invariance with
four parameters for this sector of the theory, the symmetry breaking scale $f$, the couplings
to photons and gluons, $c_{\gamma\gamma}$ and $c_{gg}$, and the dilaton self coupling $\lambda$.

A second important problem in particle physics is the nature of dark matter (DM).  Any complete
model should accomodate DM.  Thus, it is possible that these two problems, EWSB and DM, are
intertwined and the incorporation of a solution for one will impact the understanding of the 
other.  We thus extend the model of a Higgs-like dilaton with the simplest DM candidate, a 
scalar singlet which introduces two additional parameters to our model, the DM mass $m_S$ and
the DM self coupling $\lambda_S$, although this last parameter remains unconstrained by
current experimental measurements.

In the following section we review the motivation for the Higgs-like dilaton and write down 
the Lagrangian for our model.  In Section III we study the constraints that the Higgs properties
puts on the parameters of the model using a Markov Chain Monte Carlo fit using the most
recent ATLAS CMS combined results for the $\sqrt{s}=7$ and 8~TeV data sets \cite{Khachatryan:2016vau}.  In Section IV
we study constraints on the dark matter sector of our model using the measured relic abundance and
direct detection cross section limits. In section V we further constrain the model
using the recent limits on the Higgs invisible width and comment on the dilaton self 
coupling.  Finally in Section VI 
we summarize our main results and draw conclusions.

\vfill

%%%%%%%%%%%%%%%%%%%%%%%%%%%%%%%%%%%%%%%%%%%%%%%%%%%%%%%%%%%

\section{A Model of Conformal Symmetry Breaking with a Scalar Singlet Dark Matter Candidate}

\subsection{Naturalness of the Standard Model}

We begin with the observation that at the classical level the Standard Model is scale invariant except 
for the Higgs mass parameter and a soft mechanism for breaking scale invariance would 
generate the Higgs mass parameter naturally at the electroweak scale in analogy with 
supersymmetry. The scale invariance is broken by quantum corrections, that is, 
by the running of coupling constants. To see this, we can write a general Lagrangian~\cite{Coleman}
\begin{equation}
{\cal L}=\sum_{i}g_{i}(\mu){\cal O}_{i}(x)\label{eq:Eq1}
\end{equation}
where $g_{i}$ is some coupling constant defined at energy scale $\mu$
and ${\cal O}_{i}$ is an operator of dimension $d_{i}$. 
Under scale transformations $x^{\nu}\to e^{\lambda}x^{\nu}$ we obtain the following transformations:
\begin{eqnarray}
{\cal O}_{i} &\to& e^{\alpha d_{i}}{\cal O}_{i}(e^{\alpha}x), \nonumber \\
\mu&\to& e^{-\alpha}\mu \nonumber.
\end{eqnarray}
The variation of the Lagrangian under this transformation is given by
\begin{equation}
\delta {\cal L}=\sum_{i}g_{i}(\mu)(d_{i}+x^{\nu}\partial_{\nu}){\cal O}(x)_{i}+\sum_{i}\beta_{i}(g)\frac{\partial}{\partial g_{i}}{\cal L}
\end{equation}
where $\beta_i$ are the $\beta$ functions of the underlying theory.
The above implies that if the dimension of the operator $d_{i} = 4$ and the running is identically zero the theory is scale invariant. From this one obtains the divergence of the scale current $S^{\mu}=T^{\mu}_{\nu}x^{\nu}$ where
\begin{equation}
\partial_{\nu}S^{\nu}=T^{\nu}_{\nu}=\sum_{i}g_{i}(\mu)(d_{i}+x^{\nu}\partial_{\nu}){\cal O}(x)_{i}+\sum_{i}\beta_{i}(g)\frac{\partial}{\partial g_{i}}{\cal L}.
\end{equation}

Since the beta functions vanish at lowest order in perturbation theory, they cannot be responsible for the quadratic divergences in the Higgs mass parameter. This is basically the statement that the quadratic divergences are unrelated to the running of coupling constants and represent a separate explicit source of scale symmetry breaking. However, it has been noted that the appearance of quadratic divergences in a scale free theory is an artifact of the method used to regularize the loop calculations. The explicit appearance of the Higgs mass parameter in the scalar potential of the SM leads to the following trace of the energy momentum tensor
\begin{eqnarray}
T^{\mu}_{\mu,\text{tree}}&=&2\mu^{2}H^{\dagger}H \nonumber \\
T^{\mu}_{\mu,\text{one-loop}}&=&2\Delta\mu^{2}H^{\dagger}H+\sum_{i}\beta_{i}(g){\cal O}(x)_{i},
\end{eqnarray}
where $H$ are the SM Higgs fields and $\mu^2$ the bare Higgs mass parameter
with $\Delta\mu^{2}\sim\mu^{2}$~\cite{Bardeen:1995kv}. It is important to emphasize that 
the fine-tuning problem may reappear in models where the SM is the low energy description of 
a more complex theory at high energies and the sensitivity of the Higgs mass parameter to 
these new scales will depend on the scale invariance properties of these theories. 

The recent discovery of a Higgs-like resonance has disfavored technicolor/Higgsless 
models~\cite{Susskind:1978ms,Weinberg:1975gm,Weinberg:1979bn}, where the electroweak 
symmetry is broken by strong dynamics. However, there is a well motivated scenario where 
models of strong dynamics break the electroweak symmetry and yield a light resonance. 
This observation is due to the fact that the SM is scale invariant in the limit where 
the Higgs mass parameter goes to zero; and the minimum of the scalar potential has a 
flat direction, where the $vev$ will spontaneously break the approximate conformal 
invariance and the electroweak symmetry. In this scenario the Higgs is identified with 
the massless dilaton with a conformal breaking scale of $f=246$ GeV. Therefore, if the 
strongly interacting ultraviolet (UV) theory was also conformal, the condensate breaking the electroweak 
symmetry would also spontaneously break the conformal symmetry and the dilaton would have 
properties similar to the Higgs boson~\cite{Goldberger:2008zz,Bellazzini:2012vz,Vecchi:2010gj,Chacko:2012sy,Chacko:2012vm}. It is also possible to study the phenomenology a light dilaton in the presence of a fundamental scalar and this has led to recent work focusing on Higgs-dilaton mixing and how this scenario can be probed at the LHC~\cite{Abe:2012eu,Cao:2013cfa,Jung:2014zga}; however, we do not consider that here.

%%%%%%%%%%%%%%%%%%%%%%%%%%%%%%%%%%%%%%%%%%%%%%%%%%%%%%%%%%%
\subsection{Non-linearly Realized Scale Invariance}
If we assume that scale invariance is broken spontaneously by the vev of a dimensionful 
operator $\left<{\cal O}\right>=f^{n}$, where $n$ is the classical dimension of the 
operator ${\cal O}$, the spontaneous breaking of the scale invariance will imply the 
existence of a Goldstone boson, the dilaton. The dilaton field transforms non-linearly as~\cite{Coleman} 
\begin{equation}
\sigma(x)\to\sigma(e^{\alpha}x)+\alpha f,
\end{equation}

Given Eq.~(\ref{eq:Eq1}), we may incorporate non-linearly realized scale invariance by 
adding a field $\chi=e^{\sigma/f}$ serving as a conformal compensator with the transformation law
\begin{equation}
\chi(x)\to e^{\alpha}\chi(e^{\alpha}x),
\end{equation}
and make the replacement
\begin{equation}
g_{i}(\mu)\to g_{i}\left(\mu\frac{\chi}{f}\right)\left(\frac{\chi}{f}\right)^{4-d_{i}}.
\end{equation}
In the above equation $f$ is the order parameter for scale symmetry breaking.

This procedure does not guarantee that once quantum corrections are taken into account the symmetry stays intact. 
Sometimes quantum anomalies appear revealing non-invariance of the system. The system is not consistent with the 
regulator in these cases since they might introduce mass scales. Therefore, one must find a regulator 
that preserves the scale invariance such that the quantum system stays scale invariant.

There is one important difference between the theory of a spontaneously broken global symmetry and that of conformal symmetry. The latter allows for a non-derivative term in the potential
\begin{equation}
V(\chi)=\frac{\kappa}{4!}\chi^{4}.
\end{equation}
This term yields a preferred value of $f$, even in the absence of explicit sources of conformal symmetry breaking, that is the flat direction is lost and one must tune $\kappa\to 0$ such that $\left<\chi\right>=f$ remains undetermined. This issue is relevant since it introduces a fine tuning into the framework of spontaneously broken scale invariance. One way to address this issue is to introduce explicit breaking sources of scale invariance, that is an operator ${\cal O}(x)$ of scaling dimension $\Delta_{{\cal O}}$. In this case the theory is now given by
\begin{equation}
{\cal L}={\cal L}_{CFT}+\lambda_{{\cal O}} {\cal O}(x)\label{Eq:CFTB}.
\end{equation}
The work in~\cite{Bellazzini:2012vz,Chacko:2012sy} show that if the operator ${\cal O}(x)$ responsible 
for the breaking of the conformal symmetry is near marginal at the breaking scale, the dilaton 
can naturally be light and below the scale of the strong dynamics, albeit with some mild fine tuning. 
In this class of models, the mass of the dilaton is given by $m_{\sigma}\sim\sqrt{4-\Delta_{{\cal O}}}$ and 
requires that $\lambda_{{\cal O}} \ll 1$, a condition that is not expected to be satisfied in the scenarios 
of interest for EWSB. However, as stated in~\cite{Bellazzini:2012vz}, it is crucial to have a 
weakly coupled flat direction available in the theory which is hard to imagine unless the theory 
is supersymmetric or in the Goldberger-Wise stabilized Randall Sundrum model~\cite{Goldberger:1999uk}.

%%%%%%%%%%%%%%%%%%%%%%%%%%%%%%%%%%%%%%%%%%%%%%%%%%%%%%%%%%%
\subsection{The Dilaton Like Higgs with a Scalar Singlet}
Models of Technicolor are those for which the electroweak symmetry is broken dynamically 
by a strongly interacting sector, and where there is no light Higgs. The strongly interacting 
sector is assumed to be conformal in the far  UV. However, the conformal invariance is broken 
by an operator ${\cal O}$ which grows large close to the $\text{TeV}$ scale triggering EWSB. 
Within this class of theories if the exact conformal symmetry is broken spontaneously, the 
low energy effective theory contains a massless scalar, the dilaton associated with the 
breaking of the conformal symmetry~\cite{Salam:1970qk,Isham:1970gz,Isham:1971dv,Ellis:1970yd,Ellis:1971sa}.

This class of theories are expected to explicitly violate the conformal invariance since 
it is necessary to incorporate operators in the conformal sector that become strong in 
order to induce EWSB, leading to a very massive dilaton. However, in the previous section 
we mentioned that if these operators are near marginal at the breaking scale, the dilaton 
can be naturally light and lie below the scale of strong dynamics.

In this work, we are interested in theories where the SM gauge fields are not part of 
the conformal field theory (CFT), 
but gauge interactions do constitute a small source of explicit conformal symmetry 
breaking, since the CFT must be charged under the electroweak group and might also be 
charged under color. Furthermore, within this framework, the SM fermions may be elementary 
or composites of the CFT. A well motivated scenario is one where only the right-handed 
top and the Goldstone bosons needed for EWSB are composites. In addition, we incorporate a 
scalar dark matter particle, $S$, singlet under the SM gauge group and a composite of the 
CFT with no direct couplings to the SM. The stability of the dark matter particle 
is guaranteed by a $Z_{2}$ parity under which $S$ is odd while all other particles are even. 
Our framework is similar to the one studied in~\cite{Bai:2009ms,Blum:2014jca}, except that we do not 
incorporate an elementary Higgs boson candidate and we examine the possibility that the 
dilaton is the recently discovered spin $0$ particle with a mass of $125$ GeV. 

In our model, to find the couplings of the dilaton to the SM fields and the scalar dark matter 
candidate we follow the procedure of non-linearly realized scale invariance.  We start with the potential
\begin{widetext}
\begin{eqnarray}
V(\bar{\chi},S) &\approx & \frac{1}{2}m^{2}\bar{\chi}^{2} - \frac{\lambda}{3!} \frac{m^2}{f} \bar{\chi}^3
+ \frac{\bar{\chi}}{f} \sum_i (1+\gamma_{\psi_i}) m_{\psi_i}\bar{\psi}_i \psi_i
+ \left( { \frac{2\bar{\chi}}{f} + \frac{\bar{\chi}^2}{f^2} }\right)
\left[m^{2}_{W}W^{+\mu}W^{-}_{\mu}+\frac{1}{2}m^{2}_{Z}Z^{\mu}Z_{\mu}\right] \nonumber \\
& & \quad +  \frac{\alpha_{\text{EM}}}{8\pi f}c_{\gamma\gamma}\sigma F_{\mu\nu}F^{\mu\nu} 
+ \frac{\alpha_{\text{S}}}{8\pi f}c_{gg}\sigma G_{\mu\nu}G^{\mu\nu}  
+ \frac{1}{2}\bar{m}^2_S S^2 + \lambda_S S^4
\end{eqnarray}
where the dilaton $(\sigma)$ is parametrized with the non-linear realization 
as $\chi=f e^{\sigma/f}$, and we expand the dilaton about its 
VEV as $\bar{\chi}=\chi-f$,  
and $\bar{m}^2_S=m_S^2 e^{2\sigma/f}$.  We expand the exponentials to leading orders and obtain
the following parametrization for the potential:
%

%\begin{widetext}
\begin{eqnarray}
V(\sigma,S)&\approx&\frac{1}{2}m^{2}_{\sigma}\sigma^{2}+\epsilon\frac{m^{2}_{\sigma}}{v}\left(\frac{1}{2}
-\frac{1}{6}\lambda\right)\sigma^{3}
+\frac{\sigma}{f}\sum_{i}(1+\gamma_{\psi_{i}})m_{\psi_{i}}\bar{\psi}_{i}\psi_{i} 
+ \left(\frac{2\sigma}{f}+\frac{\sigma^{2}}{f^{2}}\right)\left[m^{2}_{W}W^{+\mu}W^{-}_{\mu}
+\frac{1}{2}m^{2}_{Z}Z^{\mu}Z_{\mu}\right]   \nonumber \\
& & \quad + 
\frac{\alpha_{\text{EM}}}{8\pi f}c_{\gamma\gamma}\sigma F_{\mu\nu}F^{\mu\nu} 
+ \frac{\alpha_{\text{S}}}{8\pi f}c_{gg}\sigma G_{\mu\nu}G^{\mu\nu}  
+ \frac{1}{2}m^{2}_{S}\left(1+2\epsilon\frac{\sigma}{v}
+2\epsilon^{2}\frac{\sigma^{2}}{v^{2}}\right)S^{2}+\lambda_{S}S^{4},\label{eq:modLag}
\end{eqnarray}
\end{widetext}
where $f$ denotes the scale where the conformal symmetry is spontaneously broken, $v$, 
the EWSB scale and $\epsilon=v/f$. The parameter $\lambda$ parametrizes explicit conformal 
symmetry violating effects. In addition, the mass of the dilaton and its self-interactions 
were obtained assuming that either the operator explicitly breaking the conformal symmetry 
in eqn.~\ref{Eq:CFTB} is near marginal or $\lambda_{{\cal O}}\ll1$. As 
outlined in~\cite{Goldberger:2008zz}, for a single operator with dimension $\le4$, 
this leads to a bound on the dilaton cubic coupling $\lambda<5$ which can be relaxed 
with a more elaborate conformal symmetry breaking sector. We assume that $f\ge v$ since 
we want the CFT to induce EWSB. The parameters $c_{\gamma\gamma}$ and $c_{gg}$ can be written as
\begin{equation}
c_{\gamma\gamma}=b^{\text{EM}}_{\text{IR}}-b^{\text{EM}}_{\text{UV}},~~~~~c_{gg}=b^{\text{S}}_{\text{IR}}-b^{\text{S}}_{\text{UV}}.\label{eq:Potential}
\end{equation}
The coefficients denoted by $b_{\text{IR}}$ parametrize the breaking of the conformal 
symmetry at the quantum level below the breaking scale $f$. The running in the UV, $b_{\text{UV}}$, 
does constitute an explicit source of conformal symmetry breaking and since the gluon and photon 
are elementary, there is no constraint on $b_{\text{UV}}$. The structure of the $b_{UV}$'s 
depends on the details of the CFT and 
receives contributions from both elementary and CFT states. In practice 
we have subsumed the SM loop factors \cite{Gunion:1989we} into $c_{gg}$ and $c_{\gamma\gamma}$ 
\cite{Bellazzini:2012vz,Blum:2014jca}.

In the fermion sector, the leading source of conformal symmetry breaking are the fermion 
masses and the coupling of the dilaton to fermions depend on whether the latter are elementary 
fields or composites of the CFT. This choice has been parametrized with the anomalous dimension, 
$\gamma_{i}$ in Eq. (\ref{eq:Potential}), which measures the explicit breaking of conformal 
symmetry that arises from the mixing of elementary and composite fields. Within our framework, 
we allow for the possibility that the right-handed top quark is a composite of the CFT and 
let $\gamma_{t_{R}}\to0$. In the case of elementary fermions, in theories of conformal 
technicolor, fermion masses arise from couplings of fermions to a scalar operator with the 
quantum numbers of the Higgs~\cite{Luty:2004ye,Chacko:2012vm}. In our work we will use the 
parametrization introduced in~\cite{Chacko:2012vm}, where the coupling of the dilaton to the 
top quark is given by
\begin{equation}
\frac{m_{t}}{f}(1+\delta)\sigma Q_{3}t^{c},
\end{equation}
where $\delta=\Delta_{H}-1$, and $\Delta_{H}$ is the scaling dimension of the scalar operator in the CFT.

\begin{figure*}
\centering
\begin{minipage}[t]{14cm}
\resizebox{\columnwidth}{!}{\input{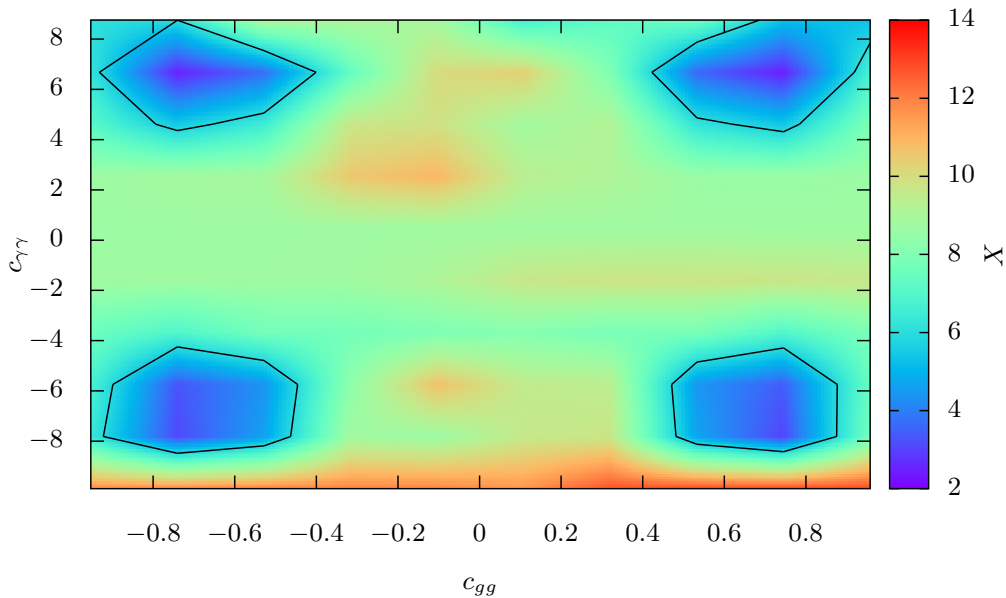}}
\end{minipage}
\caption{ Results of the MCMC fit with simulated annealing. 
The areas enclosed by the black contour lines correspond to the regions of parameter space 
for $\epsilon\ge0.98$ allowed at $95\%$ confidence level corresponding to $X < 5.9$. }
\label{fig:Fig1}
\end{figure*}

%%%%%%%%%%%%%%%%%%%%%%%%%%%%%%%%%%%%%%%%%%%%%%%%%%%%%%%%%%%
\section{Fit to Higgs Signal Strengths Using Markov Chain Monte Carlo}

In this section we present the results of a general fit to the most up-to-date 
Higgs data using the 10 Higgs production and decay signal strengths
from the most recent ATLAS CMS combined fit \cite{Khachatryan:2016vau}. The signal strengths
are defined as:
\begin{equation}
\mu_i = {\sigma_i \over {(\sigma_i)_{SM}}} \hbox{   and   } \mu^f= {{BR^f}\over{(BR^f)_{SM}}}
\end{equation}
for $\sigma_i$ with $i=ggF, \; VBF, \; WH, \; ZH, \; ttH$ 
and $BR^f$ with $f=ZZ, \; WW, \; \gamma\gamma, \; \tau\tau, \; bb$. 
For the $BR$'s one needs to also evaluate the modified total width where we rescale the 
SM Higgs partial widths by $\epsilon^2$ with the additional rescaling of the $c_{gg}$ and 
$c_{\gamma\gamma}$ factors (see for example Ref.~\cite{Coleppa:2011zx})
\begin{eqnarray}
\Gamma_{tot}& =& \epsilon^2 \left( { \sum_f \Gamma_f^{SM}  + \Gamma_{WW}^{SM} + \Gamma_{ZZ}^{SM}
+\Gamma_{\gamma Z}^{SM} }\right. \nonumber \\
& + & \left. { \left( {{c_{gg}}\over {c_{gg}^{SM} }} \right)^2 \Gamma_{gg}^{SM}
+ \left( {{c_{\gamma\gamma}}\over {c_{\gamma\gamma}^{SM} }} \right)^2 
\Gamma_{\gamma\gamma}^{SM} } \right)  +\Gamma_{inv}
\label{eqn:width}
\end{eqnarray}
where $\Gamma_i^{SM}$ are the SM Higgs partial widths  
and $\Gamma_{inv}$ is the dilaton's invisible width when decays to $SS$ are kinematically
allowed.
 For the convenience of 
the reader we quote the numerical values for the signal strengths in Table~\ref{tab:sigstrength}.
It should be noted that there are assumptions built into this set of signal strengths in that
the production signal strengths assume SM BR's and the final state signal strengths assume
SM production cross sections.  
In addition to the signal strengths we
include  an additional observable in the fit
which is the bound derived in~\cite{Barger:2012hv} on the total width of the Higgs 
boson of $6.1^{+7.7}_{-2.9}$ MeV. This result is in agreement with a CMS study that places 
an upper bound of $22$ MeV at $95\%$ CL~\cite{Khachatryan:2014iha}.

\begin{table}
\caption{ Measured signal strengths and their total uncertainties for different Higgs boson
production processes and decay channels.  The results are for the combined ATLAS CMS fits for
the combined $\sqrt{s}=7$ and 8~TeV data from ref.~\cite{Khachatryan:2016vau}.  
%The final column gives the parameter dependence of each of the signal strengths.
\label{tab:sigstrength}}
\begin{ruledtabular}
\begin{tabular}{cc}
Process			&	Value			 		\\
\hline \hline
$\mu_{ggF}$		& $1.03^{+0.16}_{-0.14}$		\\
$\mu_{VBF}$		& $1.18^{+0.25}_{-0.23}$		\\
$\mu_{WH}$		& $0.89^{+0.40}_{-0.38}$		\\
$\mu_{ZH}$		& $0.79^{+0.38}_{-0.36}$		\\
$\mu_{tth}$		& $2.3^{+0.7}_{-0.6}$			\\
$\mu^{\gamma\gamma}$ & $1.14^{+0.19}_{-0.18}$	\\
$\mu^{ZZ}$ & $1.29^{+0.26}_{-0.23}$ 	\\
$\mu^{WW}$ & $1.09^{+0.18}_{-0.16}$ 	\\
$\mu^{\tau\tau}$ & $1.11^{+0.24}_{-0.22}$  	\\
$\mu^{bb}$ & $0.70^{+0.29}_{-0.27}$  	\\
\end{tabular}
\end{ruledtabular}
%\label{tab:sigstrength}
\end{table}

To scan the parameter space we use a Markov Chain Monte Carlo (MCMC) and implement the 
Metropolis-Hastings algorithm with simulated annealing~\cite{Baltz:2006fm}. 
The target distribution is a Gaussian likelihood proposal given by
\begin{equation}
{\cal L}=\left([x_{i}]\right)=\prod_{i}exp\left[-\frac{(x_{i}-\bar{x}_{i})^{2}}{2\sigma^{2}_{i}}\right]
\label{eqn:expa}
\end{equation}
where the $\bar{x}_{i}$ are the experimental observables, and $\sigma_{i}$ their corresponding 
uncertainties. The parameters allowed to vary in the fit are those characterizing the coupling 
of the dilaton to the massless gauge bosons given below and introduced in the previous section, 
the ratio $\epsilon$  of the EWB scale $v=246$ GeV to the scale of spontaneous breaking of scale invariance, 
$f$, and the mass of the dark matter candidate $m_{S}$:
\begin{eqnarray}
\epsilon^{2} &=& \left(v/f \right)^{2}, \nonumber \\
c_{gg}&=&(b^{\text{S}}_{\text{IR}}-b^{S}_{\text{UV}}), \nonumber \\
c_{\gamma\gamma}&=&(b^{\text{EM}}_{\text{IR}}-b^{\text{EM}}_{\text{UV}}),
\label{eqn:X}
\end{eqnarray}
where the expressions for $c_{gg}$ and $c_{\gamma\gamma}$ are to leading order and to which
we should add the SM loop contributions.  In practice we let them float in the fits.
In Figure~\ref{fig:Fig1} we show the results of the MCMC in the $c_{gg}-c_{\gamma\gamma}$ plane 
where the fit has been restricted to values of $\epsilon\ge0.98$ based on the most up-to-date 
electroweak precision data (EWPD)
\cite{Ciuchini:2013pca}. Regions allowed at 95$\%$ confidence level (CL) are those for which $X <  5.9$
where $X$ is the sum of the terms in the exponential of eqn.~\ref{eqn:expa}. 
In our fit all Yukawa anomalous dimensions have been set to zero, and any information on the 
IR running of both $\beta$ functions has been diluted within the factor $c_{gg}$ and $c_{\gamma\gamma}$. 
We find that our results are consistent with results obtained in a similar analysis 
but with the $7$ TeV data set~\cite{Bellazzini:2012vz,Chacko:2012vm,Serra:2013kga}.

%%%%%%%%%%%%%%%%%%%%%%%%%%%%%%%%%%%%%%%%%%%%%%%%%%%%%%%%%%%%
\section{Dark Matter}

The introduction of the scalar singlet dark matter candidate introduces two additional parameters,
the mass of the DM candidate $m_S$ and DM self coupling $\lambda_S$ where the dilaton DM couplings 
are proportional to $m_S^2$.  In addition the dilaton  trilinear coupling comes into play through
DM self-annihilation to dilaton final states which contributes to the DM relic abundance.  $\lambda_S$
does not enter any of the expressions for the observables we are studying so it remains
unconstrained.  In 
the following subsections we explore the further contraints that the DM relic abundance and direct detection 
limits put on the parameters of the model. 

\subsection{Relic Abundance}

\begin{figure}[t]
\centering
\includegraphics[width=7.0cm]{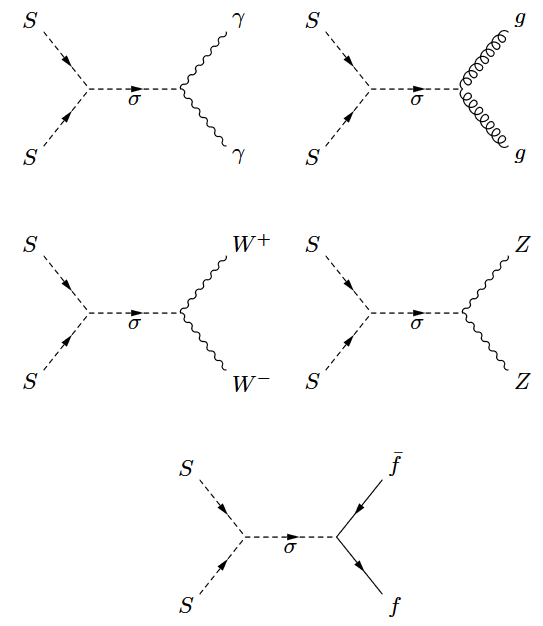}
\caption{Feynman diagrams for $SS$ annihilation into SM particles: $\gamma\gamma$, $gg$
$W^+W^-$, $ZZ$, $f\bar{f}$.}\label{fig:Fig2}
 \end{figure}

The scalar singlet dark matter candidate $S$ is stabilized by the existence of 
an unbroken $Z_{2}$ symmetry
and the interactions introduced in eqn.~\ref{eq:modLag} yield a mechanism for its 
relic abundance. The $SS$ annihilations into SM states proceed through $s$-channel dilaton exchange. 
The Feynman diagrams for these annihilation processes are shown in 
 Figure~\ref{fig:Fig2}. $SS$ annihilation into a pair 
of dilatons can proceed via $s,t$- and $u$-channels in addition to the four-point interaction
and the Feynman diagrams for these channels are shown in Figure~\ref{fig:Fig3}.
For completeness we give the expressions for these processes in Appendix~\ref{app:a}.

The present day relic abundance is determined by the DM self-annihilation in relation to the 
expansion of the universe. When the expansion dominates over the annihilation rate and the 
universe cools to a temperature below $m_{S}$, the interaction among DM particles is less 
efficient and the density freezes out. The freeze-out temperature, $T_{FO}$, at which the 
particles depart from equilibrium can be found by solving numerically the Boltzmann equation 
for the comoving particle density~\cite{Kolb:1990vq}
\begin{equation}
\frac{dn_{S}}{dt}+3Hn_{S}=-\left<\sigma v\right>(n^{2}_{S}-n^{2}_{S,eq}),
\end{equation}
where $H$ is the Hubble rate and $\left<\sigma v\right>$ is the thermally averaged annihilation 
cross section. The result is given approximately by
\begin{equation}
x_{FO}\equiv \frac{m_{S}}{T_{FO}}\approx ln\left(0.038g_{S}\frac{m_{S}M_{Pl}\left<\sigma v\right>}{g^{1/2}_{*}x^{1/2}_{FO}}\right),
\end{equation}
where $g_{*}$ is the number of relativistic degrees of freedom at the freeze out temperature, $g_S$ 
is the number of degrees of freedom of the dark matter candidate which is equal to one 
for a real scalar singlet 
and $M_{Pl}$ is the Planck mass.
The present day relic abundance is then given by
\begin{equation}
\Omega_{S}h^{2}\approx1.65\times 10^{-10}\left(\frac{\text{GeV}^{2}}{\left<\sigma v\right>}\right)\log\left(0.038g_{S}\frac{m_{S}M_{Pl}\left<\sigma v\right>}{g^{1/2}_{*}}\right).
\end{equation}
The observed relic abundance can be achieved with a value of $\left<\sigma v\right>\approx3\times 10^{-26}$ cm$^{3}$/s.

The thermally average cross section at temperature $T$ can be calculated from the 
annihilation cross section $\sigma(s)$ after summing over all the kinematically allowed 
diagrams given in Figures~\ref{fig:Fig2} and~\ref{fig:Fig3}. The result is given by
\begin{equation}
\left<\sigma v\right>=\int^{\infty}_{4m^{2}_{S}}ds\frac{(s-4m^{2}_{S})s^{1/2}K_{1}(s^{1/2}/T)}{8m^{4}_{S}TK^{2}_{2}(m_{S})/T}\sigma(s),
\end{equation}
where $K_{1}(z)$ and $K_{2}(z)$ are the modified Bessel functions of the second kind. 

Our dark matter analysis compliments the analysis by the authors in~\cite{Blum:2014jca} 
in regions of parameter space where $v\approx f$. However in our scenario, annihilations into 
fundamental scalars are absent and the value of the dilaton mass has been fixed to $125$ 
GeV and plays the role of the Higgs boson. The parameter space of interest is that 
which is compatible with the fit introduced in the previous section and with $\epsilon=0.98$.
Furthermore we do not require that 
$S$ saturates the observed relic abundance, $\Omega_{DM}=0.1199\pm0.0027$~\cite{Ade:2013zuv}. 
In Ref.~\cite{Blum:2014jca} two benchmark scenarios were considered: One where the entire SM 
is embedded into the CFT with a $c_{gg}\approx-8$ and $c_{\gamma\gamma}=11/3$ and the  
scenario where only the right handed top and the Goldstone bosons responsible for EWSB are 
composites of the CFT. In this latter scenario values of $c_{gg}=-1/3$ and 
$c_{\gamma\gamma}=-11/9$ are introduced. Both scenarios differ from the SM predictions 
of $c^{\text{SM}}_{gg}\simeq 0.67$ and $c^{\text{SM}}_{\gamma\gamma} \simeq -6.5$. 

\begin{figure}[t]
\centering
\includegraphics[width=7.0cm]{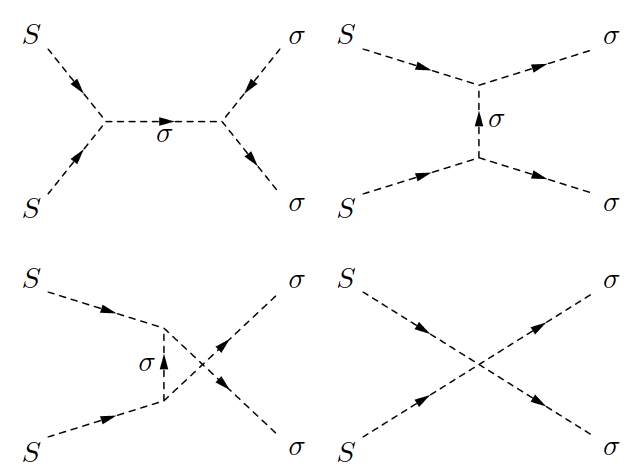}
\caption{Feynman diagrams for $SS$ annihilation into dilaton pair final states.}\label{fig:Fig3}
 \end{figure}

\begin{figure*}[t]
\centering
\includegraphics[width=10.0cm]{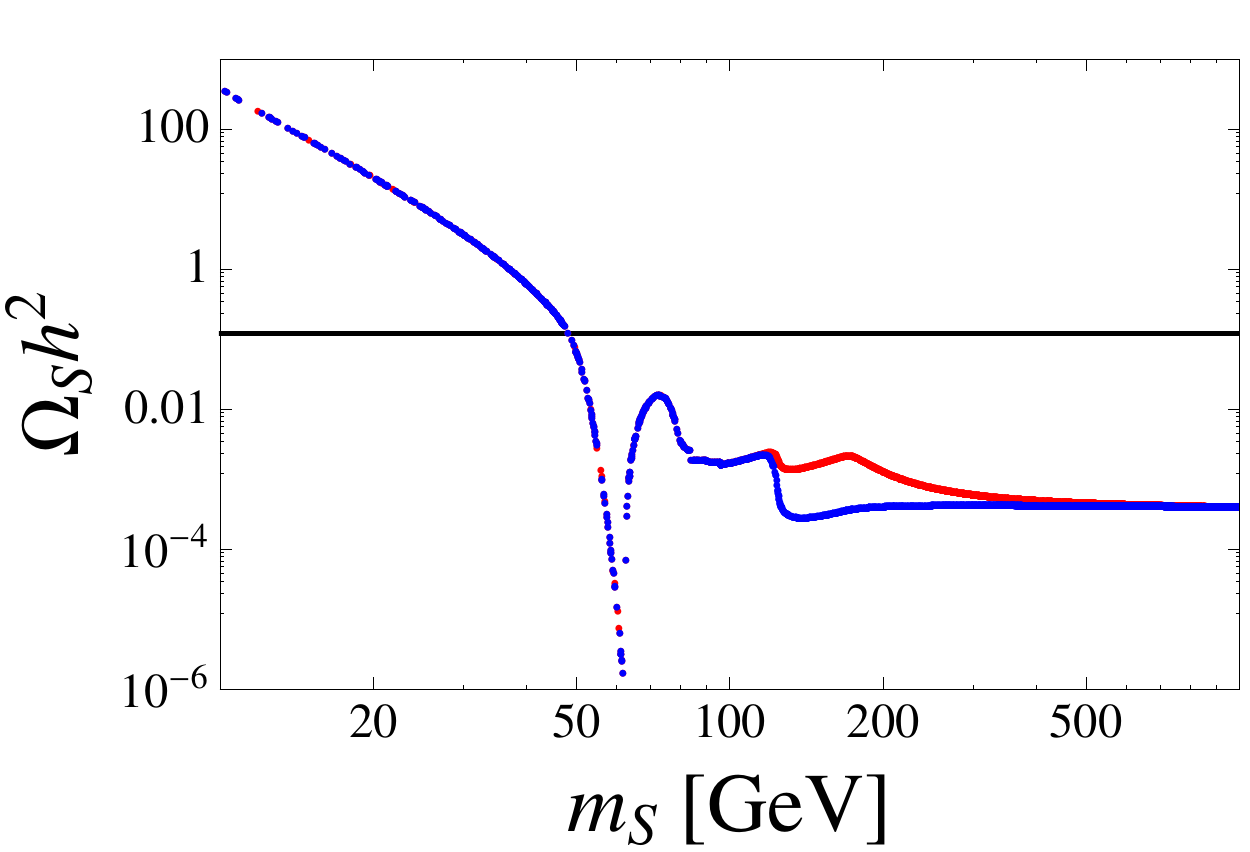}~
\caption{The present day scalar singlet DM abundance as a function of the dark matter 
mass corresponding to regions of parameter space consistent with the Higgs signal strength fit 
with the $\lambda=3$ case denoted by the red line, and the $\lambda=4\pi$ case
by the blue line (the darker, lower line when the two lines don't overlap). 
The horizontal line corresponds to the observed relic abundance, 
$\Omega_{DM}=0.1199\pm 0.0027$. }\label{fig:Fig4}
 \end{figure*}

The parameter space for $c_{gg}$ and $c_{\gamma\gamma}$ allowed at $95\%$ CL is consistent with 
the SM expectation but allows for values in the ranges $0.5\lesssim|c_{gg}|\lesssim1$ 
and $4\lesssim|c_{\gamma\gamma}|<8$. This is particularly interesting since our model 
is very similar to singlet scalar dark matter extensions of the SM~\cite{Veltman:1989vw,Silveira:1985rk,McDonald:1993ex,Burgess:2000yq,McDonald:2001vt,Barger:2007im,Goudelis:2009zz,Gonderinger:2009jp,He:2009yd,Profumo:2010kp,Yaguna:2011qn,Drozd:2011aa,Djouadi:2011aa,Kadastik:2011aa,Djouadi:2012zc,Cheung:2012xb,Damgaard:2013kva,Cline:2013gha,Baek:2014jga,Feng:2014vea,Baek:2014jga,Duerr:2015aka} 
albeit with two main differences: The mass of the dark matter candidate and its coupling to 
the Higgs-like dilaton are related, and the trilinear dilaton coupling, $\lambda$, is 
essentially a free parameter. However, the conformal algebra and unitarity imply 
 $\lambda>2$. The SM trilinear coupling corresponds to $\lambda=3$, as 
 long as $\lambda_{{\cal O}}\ll f^{2}$. This is important since annihilations into a pair of 
dilatons can be enhanced for large values of $\lambda$ and thus suppress the relic 
abundance leading to smaller direct detection cross sections as we will see in the 
next section. The study by the authors in~\cite{Goldberger:2008zz} only incorporates 
a single source of explicit conformal symmetry breaking through an operator of 
dimension $\Delta_{{\cal O}}\le4$. The bound on $\lambda$ from such an operator is $\lambda\le5$. 
In our study we will allow for $\lambda$ to be as large as $4\pi$. This has two 
implications: The annihilation cross section can be greatly enhanced as well as 
dilaton pair production at colliders. We will comment on this and also the status 
of an upcoming trilinear  Higgs coupling measurement at the LHC in the following section.

In Figure~\ref{fig:Fig4} we show the present day DM abundance as a function of the dark matter 
mass corresponding to regions of parameter space consistent with the Higgs signal strength fit  
of the previous section with the value of $\lambda=3$ denoted by the red line, and $\lambda=4\pi$ 
by the blue line (the darker lower line when the two lines don't overlap). 
The relic abundance calculation was carried out using 
MicroOMEGAs~\cite{Belanger:2013oya} with model files generated with 
FeynRules~\cite{Alloul:2013bka}.  These results are in perfect agreement with
two independent calculations using Mathematica \cite{mathematica} 
and a separate computer program using
numerical integration.  The range $10<m_{S}\lesssim50$ GeV correspond to $\sigma$-$S$ 
couplings consistent with a coupling in the Higgs-$S$ system, 
$\lambda_{p}H^{\dagger}HS^{2}$ (See for example~\cite{Duerr:2015aka}), between $0.001$ and $0.05$. 
Within this region of parameter space, the dominant annihilation channels are  
to $b\bar{b}$ and $\tau^{+}\tau^{-}$ pairs and the cross sections are suppressed by 
factors of $(m_{b,\tau}\epsilon)/v$. The observed DM abundance is achieved for a value 
of $m_{S}\sim50$ GeV and rapidly becomes greatly suppressed in the resonant region 
where $m_{S}=m_{\sigma}/2$. Beyond this point and below $m_{S}=m_{\sigma}$ 
the dominant annihilation channels are into $W^{+}W^{-}$ and $ZZ$ pairs with an enhancement 
in the abundance below the mass of the $W$ gauge boson. This region of parameter 
space corresponds to values of $\lambda_{p}$ in the range $0.05-0.3$ and annihilations 
are enhanced. We also observe that even when annihilations into a pair of dilatons are 
kinematically allowed, the value of $\lambda$ is SM-like and there is no further 
enhancement of the annihilation cross section. This situation is very different 
for $\lambda=4\pi$ where a very noticeable drop in the abundance can be seen for $m_{S}$ 
in the vicinity of $125$ GeV. Beyond $m_{S}\sim m_{t}$, annihilations into top quarks dominates 
the thermalized cross section and we see a further reduction in the relic abundance, albeit 
constant for values of $m_{S}\ge300$ GeV. In this region of parameter space $\lambda_{p}$ 
can lie between values of $1$ and $4\pi$, where the latter corresponds to dark matter masses 
above $\sim900$ GeV. Even though these values of the dark matter mass lie below the unitarity 
bound of $m_{S}\sim \sqrt{8\pi}f$~\cite{Blum:2014jca}, this region of parameter space needs 
to be rescaled by a Sommerfeld enhancement factor.

A Higgs-like dilaton augmented by a gauge singlet scalar dark matter candidate cannot 
by itself reproduce the observed abundance in most of the parameter space but it can give us an 
additional probe into the nature of the Higgs self-coupling. For masses 
$m_{S}>125$ GeV the self coupling can influence  
how important the DM annihilations into a pair of Higgs-like dilatons are and 
thus determine how large the DM abundance in that mass region can be. 

\begin{figure*}[t]
\centering
\includegraphics[width=8.0cm]{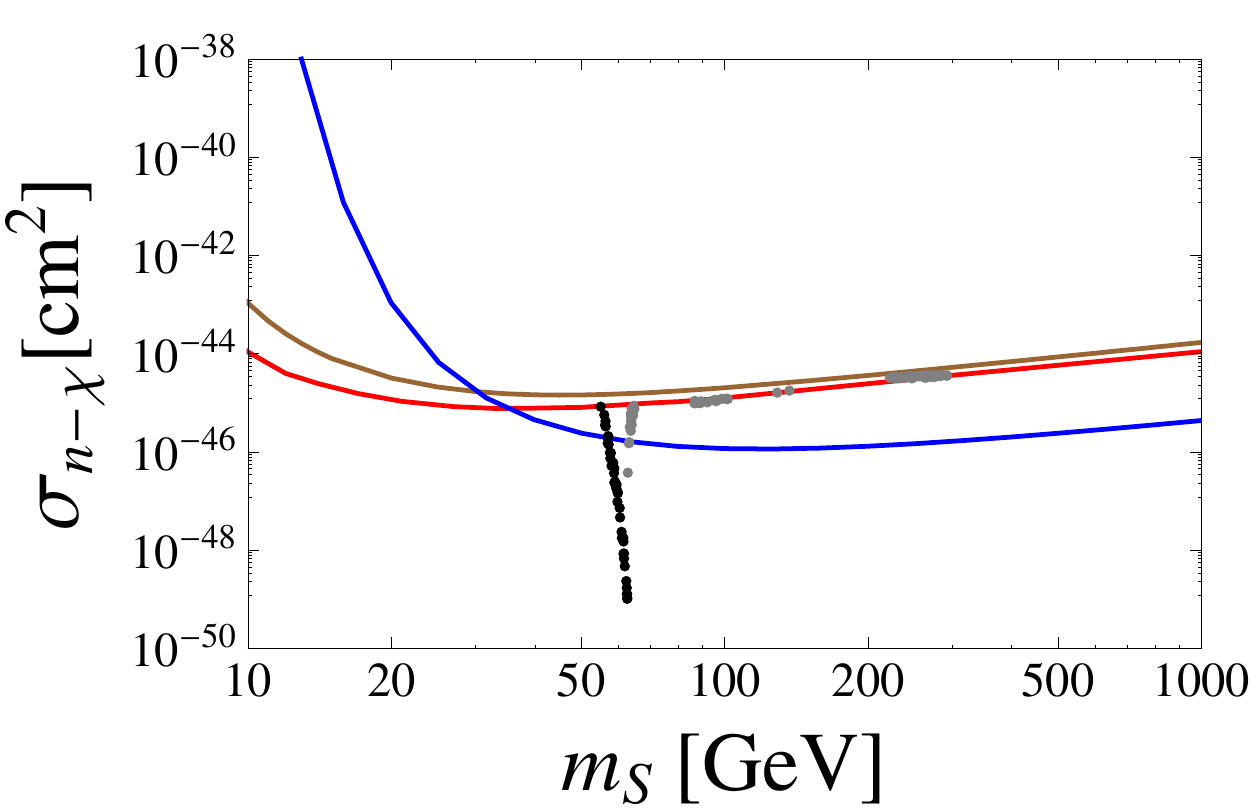}~
\includegraphics[width=8.0cm]{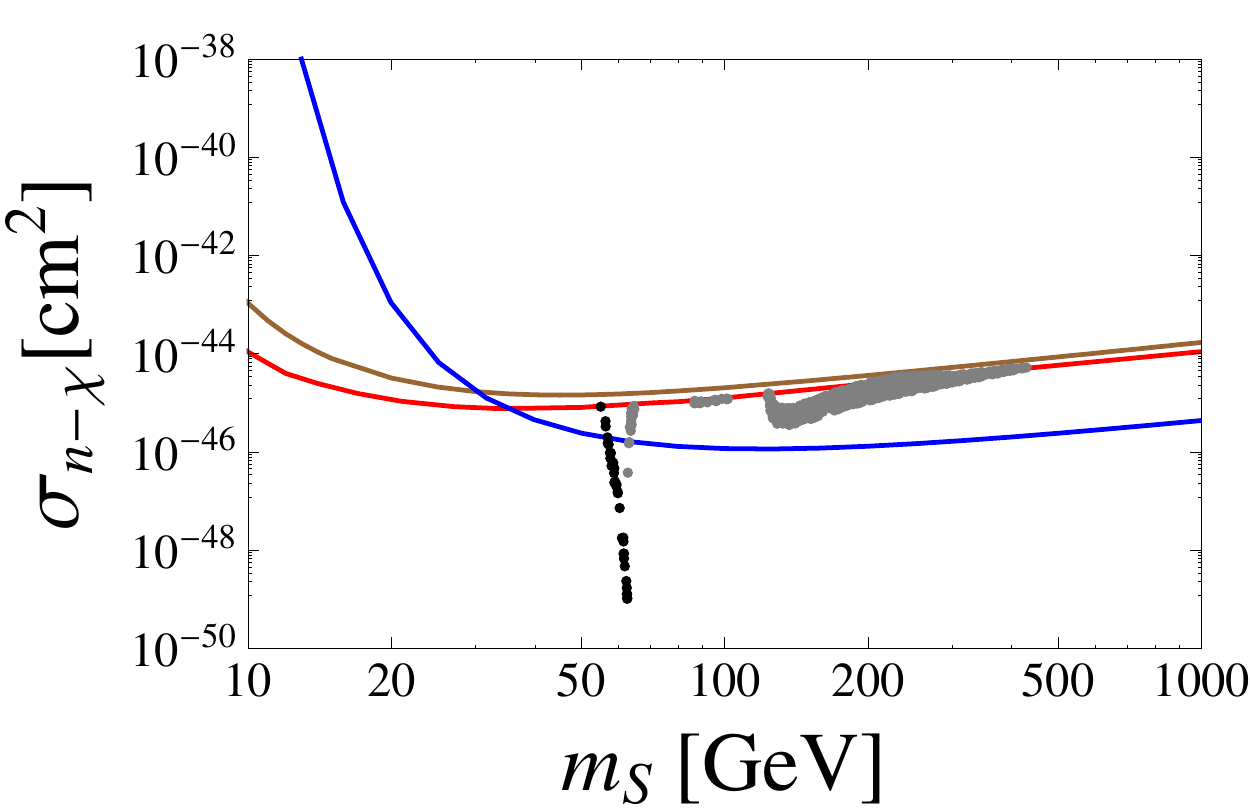}~
\caption{ Points consistent with the MCMC consistent with the observed dark matter 
relic abundance and which are consistent with direct detection limits 
form LUX~\cite{lux} (red or lower grey solid line) using a value of dilaton cubic coupling $\lambda=3$ 
(left figure) and $\lambda=4\pi$ (right figure). The points in black are those for 
which $m_{S}<m_{\sigma}/2$. We show also the projected sensitivities 
from XENON100~\cite{xenon} (brown or upper grey solid line) and DEAP~\cite{deap} (blue or black solid line). }\label{fig:fig5}
 \end{figure*}

A final comment is in order. As we saw in the previous section, the region allowed by the MCMC 
fit is symmetric in four quadrants. The SM expectation is that $c^{\text{SM}}_{gg}$ is 
positive and $c^{\text{SM}}_{\gamma\gamma}$ negative. However $c_{gg}$ and 
$c_{\gamma\gamma}$ enter quadratically into the annihilation cross section calculation 
making it insensitive to the sign of these parameters.  
We will see in the next subsection that constraints from DM direct detection measurements can further
constrain $c_{gg}$ and tell us its sign and hence rule out regions of the $c_{gg}$ parameter space.

%%%%%%%%%%%%%%%%%%%%%%%%%%%%%%%%%%%%%%%%%%%%%%%%%%%%%%%%%%%%
 \subsection{Direct Detection}

 In our model the messenger between the dark sector and the SM fields is the 
 Higgs-like dilaton so that only interactions which are spin independent between $S$ and 
 the nuclei contribute to the scattering cross section. The scattering is mediated by 
 $t$-channel exchange of a Higgs-like dilaton and it is given by the scattering cross section
 \begin{equation}
 \sigma(SN\to SN)= {1\over \pi} \left( { {{m_N m_S}\over {v m_\sigma}} } \right)^4 
 {{ \epsilon^2 f_N^2 }\over{(m_N + m_S)^2 }} \label{eqn:ddcs}
 \end{equation}
 where $m_N$ is the nucleon mass and $f_N$ is the dilaton-nucleon coupling which is 
 dependent on the interaction of the dilaton with the quarks and gluons in the nucleon
 and hence the parameters $\epsilon$ and $c_{gg}$.  To properly include the $\epsilon$ and $c_{gg}$
 dependence it is useful to refer back to the relevant pieces of our Lagrangian in eqn.~\ref{eq:modLag}:
 \begin{equation}
 {\cal L}  \supset {\epsilon \over v} \sigma \sum_i m_{\psi_i} \bar{\psi}_i \psi_i 
 +\epsilon {\alpha_s \over {8\pi v}} c_{gg} \sigma G_{\mu\nu}G^{\mu\nu}
 \end{equation}
 where in our treatment of $c_{gg}$ we have included the SM contribution.  We will 
 follow and use the approach and recent results of Cline {\it et al} \cite{Cline:2013gha} (See also~\cite{Alarcon:2011zs,Alarcon:2012nr} for a calculation of the nucleon matrix elements).
The starting point is the identity 
\begin{equation}
 f_N = \sum_i f_{q_i}.
 \end{equation}
 The contributions from $u$, $d$ and $s$ are related to the light quark matrix elements 
 and following Ref.~\cite{Cline:2013gha} we take $f_u=0.024$, $f_d=0.033$ and $f_s=0.042$. These
 values result in $f_N^{SM}=0.30$ which is in agreement with the value obtained in 
 Ref.~\cite{Cline:2013gha} which we refer to for details of the derivation.  
 The heavy quark ($c$, $b$, $t$) content of the nucleon is negligible so that 
 the contribution for heavy quarks comes from
 the triangle diagram that generates the $\sigma G_{\mu\nu} G^{\mu\nu}$ coupling which is
 dominated by the top quark contribution.  
 In our model we need to include the $\langle N | G_{\mu\nu}G^{\mu\nu} | N \rangle$
 explicitly.  Because we have subsumed the SM heavy quark contribution into $c_{gg}$ we 
 include the gluon contribution into $f_N$ by rescaling the SM gluon contribution resulting
 in the following expression for $f_N$:
 \begin{equation}
 f_N=\epsilon ( f_u + f_d + f_s +f_{G} ) = \epsilon \left( {0.099 + 0.201 {{c_{gg}}\over {c_{gg}^{SM}}} } \right).
 \label{eqn:fn}
 \end{equation}
 Using eqns.~\ref{eqn:ddcs} and \ref{eqn:fn}, the parameters allowed by the fits to the
 Higgs signal strengths consistent with the DM relic abundance,
 we obtain the direct detection cross sections shown in Fig.~\ref{fig:fig5} along with
 the LUX~\cite{lux}, XENON100~\cite{xenon} and projected DEAP~\cite{deap} direct detection limits. 
The points shown in Fig.~\ref{fig:fig5} were obtained using $c_{gg}>0$, 
 consistent with the sign of the SM value, $c^{SM}_{gg}$. 
 If we relax this constrain and allow negative values of $c_{gg}$ from the MCMC fit the 
 resulting value of $f_{N}$ will be small enough to yield cross sections that evade direct detection 
 constraints in most of the region of parameter space for which $m_{S}>50$ GeV, for both small and large $\lambda$.
 %%%%%%%%%%%%%%%%%%%%%%%%%%%%%%%%%%%%%%%%%%%%%%%%%%%%%
 \section{Other Higgs properties}
 
 %%%%%%%%%%%%%%%%%%%%%%%%%%%%%%%%%%%%%%%%%%%%%%%%%%%%%%%%%%%%
\subsection{Invisible Dilaton Width}
In the previous sections we analyzed the constraints on the dilaton, which continues to be a viable candidate for
the $125$~GeV scalar resonance observed at the LHC. We have also analyzed whether within the 
allowed parameter space, the dilaton can couple to a scalar dark matter particle that does 
not over close the universe and does not violate current dark matter direct detection 
constraints. Within our framework, the dilaton can decay to the DM candidate, $S$, and thus 
contribute to its invisible width. If the dilaton is to mimic the SM Higgs boson, its 
branching ratio to invisible final states must lie below the most recent $95\%$ CL upper bound from 
the ATLAS collaboration of $BR(H\to inv.)<0.28$~\cite{Aad:2015pla}. In this model, the 
invisible decay width of the dilaton is given by
\begin{equation}
\Gamma (\sigma\to SS)=\frac{m^{4}_{S}\epsilon^{2}}{8\pi m_{\sigma} v^{2}}\sqrt{1-4\frac{m^{2}_{S}}{m^{2}_{\sigma}}},
\end{equation}
with the branching ratio given by
\begin{equation}
BR(\sigma\to inv.)=\frac{\Gamma(\sigma\to SS)}{\Gamma(\sigma\to SM)+\Gamma(\sigma\to SS)}
\end{equation}
where $\Gamma(\sigma\to SM)$ can be read from eqn.~\ref{eqn:width}.

In our model, a scalar dark matter particle with a mass $m_{S}\approx50$ GeV can 
saturate the observed dark matter relic abundance. However, this mass corresponds to a 
$S-\sigma$ coupling that leads to a direct detection cross section above the experimental bounds. 
In our framework masses below $m_{\sigma}/2$ (black points in Figure~\ref{fig:fig5}) 
lead to an invisible branching fraction well above the experimental bound given above. 
This is consistent with the results obtained for a singlet extension of the SM where 
the Higgs-$S$ couplings, 
 $\lambda_{p} \gtrsim 0.015$ are ruled out by 
this observable. Masses in the range $10-m_{\sigma}/2$ correspond to values of $\lambda_{p}$ 
in the Higgs-$S$ system in the range $\sim0.01-0.07$.
 %%%%%%%%%%%%%%%%%%%%%%%%%%%%%%%%%%%%%%%%%%%%%%%%%%%%%%%%%%%%
 
 \subsection{The dilaton self coupling}
 Here we comment on the dilaton self coupling. The observation of what appears 
 to be a SM-like Higgs boson has given us an insight into the mechanism of EWSB, but in order 
 to fully establish the Brout-Englert-Higgs mechanism detailed knowledge of the Higgs potential 
 and precise measurement of its parameters is necessary. Within the SM, the cubic and quartic
  Higgs self-couplings are uniquely defined:
 \begin{equation}
 \lambda^{\text{SM}}_{HHH}=\lambda^{\text{SM}}_{HHHH}=\frac{m^{2}_{H}}{2v^{2}}
 \end{equation}
 
 Double and trilinear Higgs production can help probe the cubic and quartic 
 self-interactions governing the Higgs potential. Any deviation from the SM value will 
 be a sign of new physics and force us to rethink and possibly reformulate the recipe for EWSB.
  In our work, we have focused on a particular form for the dilaton potential which arises from the 
  contributions that explicitly violate conformal invariance. We have incorporated the potential 
  introduced in ref.~\cite{Goldberger:2008zz} and given in eq.~\ref{eq:modLag}. With this parametrization, 
  the cubic coupling is given by
 \begin{equation}
 \lambda_{\sigma\sigma\sigma}=\epsilon\left(\frac{m_{\sigma}}{v}\right)^{2}\left(1-\frac{1}{6}\lambda\right),
 \end{equation}
 with $\lambda$ directly related to the operator given in eq.~\ref{Eq:CFTB}:
 \begin{equation}
\lambda \approx \left\{
\begin{array}{rl}
(\Delta_{{\cal O}}+1)+{\cal O(\lambda_{{\cal O}})} & \text{when } \lambda_{{\cal O}} \ll 1,\\
5+{\cal O}\left(|\Delta_{{\cal O}}-4|\right) & \text{when } |\Delta_{{\cal O}}-4|\ll 1,
\end{array} \right.
\end{equation} 
 where $\Delta_{{\cal O}}$ is the scaling dimension of the symmetry breaking operator. 
 A SM-like coupling corresponds to a scaling dimension of $\Delta_{{\cal O}}=2$ 
 for a case where $\lambda_{{\cal O}}/f^{2}\ll1$. More elaborate ways of breaking scale 
 invariance, and in particular with near-marginal operators can lead to values of $\lambda$ 
 closer to its non-perturbative value of $4\pi$; however this may involve a certain degree of 
 fine tuning to generate a dilaton mass below the conformal symmetry breaking 
 scale~\cite{Bellazzini:2012vz,Chacko:2012sy}. For $\lambda=4\pi$, the corresponding 
 value of the cubic coupling is approximately a factor of $1.5$ larger 
 than $\lambda^{\text{SM}}_{HHH}$. Thus, we can get partial insight into the dynamics 
 leading to EWSB and whether the observed scalar is indeed the one predicted by the SM 
 or a dilaton by measuring double Higgs production at the LHC; its prospects which have 
 been actively studied and that show that a combination of channels as well as large data 
 sets are needed to overcome the SM backgrounds~\cite{Dolan:2012rv,Dolan:2012ac,Barr:2013tda,Dolan:2013rja}. 
 A detailed review that includes beyond the LHC colliders in~\cite{Panico:2015cst}, 
 shows how a $1$ TeV lepton collider such as the ILC can achieve a precision of $\sim 30\%$ 
 on the determination of the cubic self-coupling. In addition, a recent CMS study places an 
 upper bound on non-resonant double Higgs production of $200$ times the SM rate~\cite{CMS:2016ugf} 
 using center of mass energies of $\sqrt{s}=13$ TeV and $2.7$ fb$^{-1}$ of data. 
 This limit is still much weaker than the limit obtained by ATLAS with the full 
 data set at $\sqrt{s}=8$ TeV. The observed limit in this case is approximately $70$ 
 times the SM rate~\cite{Aad:2015xja}.

\section{Summary}

We have studied the phenomenology of a model that treats the physical Higgs boson of the SM
as a dilaton, the pseudo-Goldstone boson of spontaneous conformal symmetry breaking and 
introduces a scalar singlet dark matter candidate. We 
fitted the parameters of the model to the measured Higgs boson properties and found that 
the parameters of the model sensitive to these observables are 
forced to be consistent with those of the standard model to a high degree modula a sign 
ambiguity for two of them; $c_{gg}$ and $c_{\gamma\gamma}$.  Using the allowed region of the 
parameter space we then study constraints on the model from DM constraints.  We find that 
$m_S$ is ruled out for masses less than $\sim 50$~GeV by the allowed values of DM 
relic abundance.  When direct detection limits by the XENON and LUX experiments are imposed
only a small mass region remains assuming a SM-like dilaton self coupling. When one assumes
the self coupling is more like a strongly interacting composite theory a second region
survives in the mass region of $\sim 125 - 300$~GeV. In addition, the non-linear nature of the 
coupling between the dilaton and $S$ yield large couplings for masses $m_{S}<m_{\sigma}/2$ and 
thus DM candidates below this limit are ruled out by the current upper limit of the Higgs 
invisible width. The higher mass region will be probed 
in the near future by the DEAP DM search experiment. We also find that the region of 
parameter space allowed by direct detection experiments is enhanced when one allows 
for a negative value of $c_{gg}$. When the Higgs was first discovered, a slight 
enhancement in the $H\to\gamma\gamma$ channel led to work that could potentially 
explain this excess with an effective negative Higgs-gluon coupling~\cite{Reece:2012gi}, 
due to an overall negative coupling between the Higgs and top quark or new physics running 
in the loop. In this work we see that the structure of the CFT dictates the sign of $c_{gg}$, 
but we also observe that whatever this structure is, a fit to the Higgs data leads to an 
absolute value consistent with the SM. Furthermore, we observe that future precision 
measurements on the parameters governing the Higgs potential can tell us if we are 
indeed observing the scalar responsible for the Brout-Englert-Higgs EWSB mechanism or 
a dilaton. If these measurements deviate from the SM expectation, we can learn more 
about the CFT for energies $f\gtrsim246$ GeV and the mechanism responsible for a light dilaton, 
given that its natural mass will be on the order of $4\pi f$.

One important lesson to be learned from this work is that if we give up the idea of 
explaining with a single model the complete nature of the dark matter and simply work 
on a first order theory producing a fraction of the abundance with a dark matter component 
with spin independent interactions with nuclei, one can learn in a way complimentary to 
precise measurements of the Higgs scalar potential parameters the nature of the UV theory leading to EWSB.

\acknowledgments

The authors have benefited from conversations with Heather Logan, Alex Poulin and Rouzbeh
Yazdi.
This research was supported in part 
the Natural Sciences and Engineering Research Council of Canada under Grant No. 121209-2009 SAPIN. 

\appendix
\section{$SS$ Self Annihilations Cross Sections}
\label{app:a}

In this appendix we include the expressions for the $SS$ annihilation cross sections 
used to calculate the DM relic abundance.

\begin{widetext}
\begin{equation}
\sigma(SS\to ZZ)=\frac{1}{8\pi}\frac{\sqrt{s-4m_{Z}^{2}}}{\sqrt{s-4m_{S}^{2}}}\left(\frac{m_{S}\epsilon}{v}\right)^{4}\frac{s}{(s-m_{\sigma}^{2})^{2}+\Gamma_{\sigma}^{2}m_{\sigma}^{2}}\left(1+\frac{12m_{Z}^{4}}{s^{2}}-\frac{4m_{Z}^{2}}{s}\right)
\end{equation}

\begin{equation}
\sigma(SS \to W^+W^-)=\frac{1}{4\pi}\frac{\sqrt{s-4m_{W}^{2}}}{\sqrt{s-4m_{S}^{2}}}\left(\frac{m_{S}\epsilon}{v}\right)^{4}\frac{s}{(s-m_{\sigma}^{2})^{2}+\Gamma_{\sigma}^{2}m_{\sigma}^{2}}\left(1+\frac{12m_{W}^{4}}{s^{2}}-\frac{4m_{W}^{2}}{s}\right)\end{equation}

\begin{equation}
\sigma (SS \to {\gamma\gamma})=
\frac{(c_{\gamma\gamma}\alpha_{EM})^{2}}{64\pi^{3}}\frac{s}{(s-m_{\sigma}^{2})^{2}-\Gamma_{\sigma}^{2}m_{\sigma}^{2}}\left(\frac{m_{S}\epsilon}{v}\right)^{4}\left(1-\frac{4m_{S}^{2}}{s}\right)^{-\frac{1}{2}}
\end{equation}

\begin{equation}
\sigma (SS \to gg)=
\frac{(c_{gg}\alpha_{S})^{2}}{8\pi^{3}}\frac{s}{(s-m_{\sigma}^{2})^{2}-\Gamma_{\sigma}^{2}m_{\sigma}^{2}}\left(\frac{m_{S}\epsilon}{v}\right)^{4}\left(1-\frac{4m_{S}^{2}}{s}\right)^{-\frac{1}{2}}
\end{equation}

\begin{equation}
\sigma (SS \to f\bar{f})=\frac{m_{f}^{2}N_{C}}{2\pi s}\left(\frac{m_{S}\epsilon}{v}\right)^{4}\frac{1}{(s-m_{\sigma}^{2})^{2}+\Gamma_{\sigma}^{2}m_{\sigma}^{2}}\frac{(s-4m_{f}^{2})^{\frac{3}{2}}}{\sqrt{s-4m_{S}^{2}}}
\end{equation}
where $N_C=1$ for leptons and 3 for quarks.

\begin{eqnarray} 
\sigma(SS \to \sigma\sigma)  = \frac{1}{16 \pi s} 
\frac{\sqrt{s-4m_{\sigma}^{2}}}{\sqrt{s-4m_{S}^{2}}}\left(\frac{m_{S}\epsilon}{v}\right)^{4}
& &
\int_{-1}^{+1} d\cos\theta 
\left[ { \frac{m_\sigma^4 (3-\lambda)^2}{(s-m_\sigma^2)^2 + \Gamma_\sigma^2 m_\sigma^2} } \right. \nonumber \\
& & \qquad 
\frac{4 m_S^2 m_\sigma^2 (3-\lambda) (s-m_\sigma^2)}{(s-m_\sigma^2)^2 + \Gamma_\sigma^2 m_\sigma^2 } 
\left( { \frac{1}{t} + \frac{1}{u} } \right) 
+ \frac{4(3-\lambda) m_\sigma^2 (s-m_\sigma^2)}{ (s-m_\sigma^2)^2 + \Gamma_\sigma^2 m_\sigma^2 } \nonumber \\
& & \qquad \left. {
4m_s^4 \left( { \frac{1}{t} + \frac{1}{u} } \right)^2
+ 8 m_S^2 \left( { \frac{1}{t} + \frac{1}{u} } \right) +4 }\right]
\end{eqnarray}
where $s$, $t$ and $u$ are the appropriate Mandelstam variables.

\end{widetext}

%%%%%%%%%%%%%%%%%%%%%%%%%%%%%%%%%%%%%%%%%%%%%%%%%%%%%%%%%%%%

\end{document}